\documentclass[11pt,tightenlines,superscriptaddress,floatfixolumn,english,aps,notitlepage,nofootinbib]{revtex4-1}

\usepackage{amsmath}
\usepackage{esint}
\usepackage{graphicx}
\usepackage{hyperref}
\usepackage{color}
\usepackage{xcolor}
\usepackage[utf8]{inputenc} %%% needed  for some of the references
\advance\voffset -0.2in

\def\ba{\begin{eqnarray}}
\def\ea{\end{eqnarray}}

\begin{document}

\title{Wounded quarks and diquarks in heavy ion collisions}

\author{A.Bialas}
\email{bialas@th.if.uj.edu.pl} 

\author{A.Bzdak}
\email{adam.bzdak@fis.agh.edu.pl}

\affiliation{M.Smoluchowski Institute of Physics, Jagellonian University, 
Cracow, Poland} 

\begin{abstract}
A model in which the soft collisions of the nucleon are described in terms of
interactions of its two constituents (a quark and a diquark) is proposed. When
adjusted to describe precisely the elastic proton-proton scattering data and
supplemented with the idea of wounded constituents, the model accounts rather
well for the centrality dependence of particle production in the central
rapidity region at RHIC energies.
\end{abstract}

\maketitle

\section{Introduction}

The data on production of particles in relativistic heavy ion collisions
collected during the operation of the RHIC accelerator
\cite{AuAu-130,AuAu-19-200,AuAu-62} indicate that the wounded nucleon model
\cite{wnm}, does not describe correctly the observed particle multiplicities.
Contrary to the model predictions, the particle density per one wounded
nucleon in the central rapidity region (i) increases slowly with centrality of
the collision and (ii) substantially exceeds that observed in nucleon-nucleon collisions.

This is not too surprising because the model was always considered only as a
first, rough approximation. Indeed, soon after the original model was
formulated, a possible improvement was proposed in the form of the wounded
quark model \cite{bcf}. Recently, it was suggested \cite{nou} that the wounded
quark model can account for the main features of the data in the central
rapidity region. To obtain this result, however, it was necessary to assume a
rather large number of quark-quark collisions in the nucleon-nucleon
scattering which is difficult to justify. If more realistic value of this
number is used, the model predicts larger multiplicities than actually
observed \cite{bw}. Nevertheless, these results show that the idea of a
wounded constituent model may be not far from reality.

In the present paper we propose another generalization of the wounded nucleon
model, based on the idea that the nucleon is composed of two constituents: one
constituent quark forming a colour triplet and one constituent diquark forming
a colour antitriplet. We assume, furthermore, (i) that particle production
from these constituents is independent of the number of interactions they
underwent and (ii) both constituents produce the same number of particles. It
turns out that these two assumptions are sufficient to describe correctly the
PHOBOS data \cite{AuAu-130,AuAu-19-200,AuAu-62} on particle production in the
central ($y=0$) rapidity region.

In the next section we formulate the model in more detail. In Section 3 the
determination of the model parameters from proton-proton elastic scattering is
described. Particle production in Au-Au collisions is discussed and compared
with the PHOBOS data in Section 4. Our comments and conclusions are listed in
the last section.

\section{The model}

As indicated in the introduction, we assume that in the soft collisions
nucleon can be approximated by a system composed of the two constituents - a
quark and a diquark - acting independently.

For the process of particle production we assume that each constituent (quark
or diquark) which underwent at least one inelastic collision emits a certain
amount of secondary partons. This number is independent of the number of
collisions this constituent underwent afterwards. As this is obviously a
straightforward generalization of the wounded nucleon concept we call them
wounded constituents. Furthermore, we assume that the wounded constituent
quark and diquark emit secondary partons in approximately the same manner.

The arguments in favour of such an approach are collected in the last section.
Here we would like to emphasize, however, that they may apply only to
\textit{soft} collisions, i.e. those in which the transverse mass of the
emitted partons does not exceed $\sim200$ MeV. The analogous limits for the
observed hadrons are, naturally, somewhat higher.

It should be also pointed out that the proposed description can only be
justified far from the fragmentation regions of the projectile and of the
target. Therefore in this paper we restrict the comparison with the data to
the central rapidity region.

The consequence of these assumptions, fundamental for our conclusions, is that
the differential multiplicity of partons emitted in the nucleus-nucleus
inelastic collisions can be represented as:%
\begin{equation}
\frac{dN_{AB}}{dy}=w_{A}^{(q+d)}F_{+}(y)+w_{B}^{(q+d)}F_{-}(y), \label{ab}%
\end{equation}
where $w_{A}^{(q+d)}$ and $w_{B}^{(q+d)}$ are the numbers of wounded
constituents (quarks and diquarks) in nucleus $A$ and $B$, whereas $F_{+}$,
$F_{-}$ are the differential multiplicities of partons emitted by one wounded
constituent in $A$ and $B$, respectively\footnote{This formula of course is
valid for any distribution, not necessarily in rapidity.}.

When (\ref{ab}) is applied to nucleon-nucleon collisions we obtain:
\begin{equation}
\frac{dN_{NN}}{dy}=w_{N}^{(q+d)}\left[  F_{+}(y)+F_{-}(y)\right]  . \label{nn}%
\end{equation}

These two equations summarize the relation between particle production in
nucleon-nucleon and nucleus-nucleus collisions implied by the model.

For the symmetric collisions ($A=B$) we obtain the particularly simple and
elegant result:
\begin{equation}
R_{AA}\equiv\frac{dN_{AA}/dy}{dN_{NN}/dy}=\frac{w_{A}^{(q+d)}}{w_{N}^{(q+d)}}.
\label{start}%
\end{equation}
Since the R.H.S. of this equation is independent of the phase-space region
where $R_{AA}$ is measured\footnote{Provided it is far enough from the
fragmentation regions.}, but depends on the centrality of the collision, this
is indeed a very strong consequence of the model.

At the vanishing c.m. rapidity, (\ref{ab}) and (\ref{nn}) imply a simple
relation even for asymmetric collisions:
\begin{equation}
R_{AB}(y=0)\equiv\frac{dN_{AB}(y=0)/dy}{dN_{NN}(y=0)/dy}=\frac{w_{A}%
^{(q+d)}+w_{B}^{(q+d)}}{2w_{N}^{(q+d)}}. \label{start'}%
\end{equation}

To make a full use of (\ref{start}) and (\ref{start'}), it is necessary to
evaluate $w_{N}^{(q+d)}$ and $w_{A}^{(q+d)},w_{B}^{(q+d)}$ as function of the
impact parameter of the collision. The corresponding formulae are obtained by
a straightforward counting of probabilities \cite{cm,wnm}. For the average
number of wounded quarks and diquarks in each of the colliding nuclei
($w_{A}^{(q+d)}=W_{q}+W_{d}$), at a fixed impact parameter $b$, one obtains:%
\begin{equation}
W_{a}(b)=\frac{A}{\sigma_{AA}(b)}\int T(b-s)\left(  1-\left[  1-\sigma
_{aq}T_{aq}\left(  s\right)  \right]  ^{A}\left[  1-\sigma_{ad}T_{ad}\left(
s\right)  \right]  ^{A}\right)  d^{2}s, \label{www}%
\end{equation}
where $a$ denotes $q$ or $d$, $\sigma_{AA}(b)\equiv d^{2}\sigma_{AA}/d^{2}b$
is the inelastic differential $AA$ cross-section\footnote{For heavy nuclei
$\sigma_{AA}(b)=1$, except at very large impact parameters which are of no
interest even for most peripheral events measured at RHIC. In case of $AuAu$
collisions, using the optical approximation, we have verified that
$\sigma_{AuAu}(b)=1$ for $b\leq14$ fm, corresponding to $W\geq5$.}, $T(b)$ is
the nuclear thickness function (normalized to unity). $T_{ab}(b)$ is given by:%
\begin{equation}
T_{ab}(b)=\frac{1}{\sigma_{ab}}\int\sigma_{ab}(s)T(b-s)d^{2}s, \label{tab}%
\end{equation}
where $\sigma_{ab}(s)\equiv{d}^{2}{\sigma_{ab}}/{d^{2}s}$ are the differential
cross-sections of the constituents (in impact parameter plane) and
$\sigma_{ab}=\int\sigma_{ab}(s)d^{2}s$ are the corresponding total inelastic
cross-sections ($ab$ denotes $qq,qd$ or $dd$).

Note that the formulae (\ref{www}) and (\ref{tab}) take into account the
impact parameter dependence of the constituent cross-sections. If this
dependence is neglected [$\sigma_{ab}(s)=\sigma_{ab}\delta^{2}(s)$],
$T_{ab}(b)\equiv T(b)$.\footnote{We have verified that this is a poor
approximation for peripheral collisions.}

For the number of wounded constituents in nucleon-nucleon collisions we have,
similarly, $w_{N}^{(q+d)}=w_{q}+w_{d}$ with:
\begin{equation}
w_{q,d}=\frac{1}{\sigma_{NN}}\int h_{q,d}(b)d^{2}b, \label{wq+wd}%
\end{equation}
where $\sigma_{NN}$ is the total inelastic nucleon-nucleon cross-section and
$h_{a}$ is given by:%
\begin{align}
h_{a}(b)  &  =\int d^{2}s_{q}d^{2}s_{q}^{\prime}d^{2}s_{d}d^{2}s_{d}^{\prime
}D(s_{q},s_{d})D(s_{q}^{\prime},s_{d}^{\prime})\nonumber\\
&  \left\{  1-\left[  1-\sigma_{ad}(b+s_{d}^{\prime}-s_{a})\right]
[1-\sigma_{aq}(b+s_{q}^{\prime}-s_{a})]\right\}  ,
\end{align}
with $D(s_{q},s_{d})$ being the effective thickness of the nucleon.

It remains to determine the cross-sections of the constituents and their
distribution inside the nucleon. This demands a detailed analysis of the
nucleon-nucleon collisions, as discussed in the next section.

\section{Nucleon-nucleon collisions}

The distribution of the constituents inside the nucleon and their
cross-sections are not known. We propose to determine them from the analysis
of the data on proton-proton elastic scattering.

Consider first the inelastic nucleon-nucleon collisions. Following the idea
that the interaction can be described as the interaction of two independent
constituents in each of the nucleons, we have \cite{cm,Glauber}:
\begin{align}
1-\sigma(s_{q},s_{d};s_{q}^{\prime},s_{d}^{\prime};b)  &  =[1-\sigma
_{qq}(b+s_{q}^{\prime}-s_{q})][1-\sigma_{qd}(b+s_{d}^{\prime}-s_{q}%
)]\nonumber\\
&  [1-\sigma_{dq}(b+s_{q}^{\prime}-s_{d})][1-\sigma_{dd}(b+s_{d}^{\prime
}-s_{d})],
\end{align}
and%
\begin{equation}
\sigma(b)=\int d^{2}s_{q}d^{2}s_{q}^{\prime}d^{2}s_{d}d^{2}s_{d}^{\prime
}D(s_{q},s_{d})D(s_{q}^{\prime},s_{d}^{\prime})\sigma(s_{q},s_{d}%
;s_{q}^{\prime},s_{d}^{\prime};b),
\end{equation}
where $s_{q}(s_{q}^{\prime})$, $s_{d}(s_{d}^{\prime})$ are transverse
positions of the quarks and diquarks in the two colliding nucleons.

From the unitarity condition we deduce:
\begin{equation}
t_{el}(b)=1-\sqrt{1-\sigma(b)}. \label{tel}%
\end{equation}
This allows one to evaluate the elastic and total cross-sections. By comparing
them with data one can obtain information on the parameters of the model.

Since the nuclear cross-sections are not sensitive to the exact shape of the
impact parameter dependence of the constituent cross-sections (as the nuclear
radius is much larger than that of the nucleon), we parametrized $\sigma
_{ab}(s)$ using simple Gaussian forms:
\begin{equation}
\sigma_{ab}(s)=A_{ab}e^{-s^{2}/R_{ab}^{2}}. \label{pab}%
\end{equation}

The radii $R_{ab}$ were constrained by the condition $R_{ab}^{2}=R_{a}%
^{2}+R_{b}^{2}$ where $R_{a}$ denotes the quark or diquark's radius (a natural
constraint for the Gaussians).

From (\ref{pab}) we deduce the total inelastic cross sections: $\sigma
_{ab}=\pi A_{ab}R_{ab}^{2}$ and we also demand that the ratios of
cross-sections satisfy the natural condition:\footnote{We have verified that
the main results of this paper are not sensitive to the exact values of these
ratios. A detailed analysis of elastic pp scattering (and, particularly of
this assumption) will be given elsewhere \cite{ab-ab}.}%
\begin{equation}
\sigma_{qq}:\sigma_{qd}:\sigma_{dd}=1:2:4, \label{124}%
\end{equation}
expressing the idea that there are twice as many partons in the constituent
diquark than those in the constituent quark. This allows to express $A_{qd}$
and $A_{dd}$ in terms of $A_{qq}$.

For the distribution of the constituents we again take a Gaussian:
\begin{equation}
D(s_{q},s_{d})=\frac{1+\lambda^{2}}{\pi R^{2}}e^{-(s_{q}^{2}+s_{d}^{2})/R^{2}%
}\delta^{2}(s_{d}+\lambda s_{q}). \label{D}%
\end{equation}

The parameter $\lambda$ has the physical meaning of the ratio of the quark and
diquark masses and satisfies, $1/2\leq\lambda=m_{q}/m_{d}\leq1$ (the delta
function guarantees that the center-of-mass of the system moves along the
straight line).

Thus, finally, the model contains $5$ free parameters.

Using this formulation and the formula (\ref{tel}) we have evaluated the
elastic and total proton-proton cross-sections and adjusted the parameters by
demanding that (i) total inelastic cross section (ii) slope of the elastic
cross section (iii) position of the first diffractive minimum in elastic cross
section and (iv) height of the second maximum in elastic scattering are in
agreement with data.\begin{figure}[h]
\begin{center}
\includegraphics[scale=1.3]{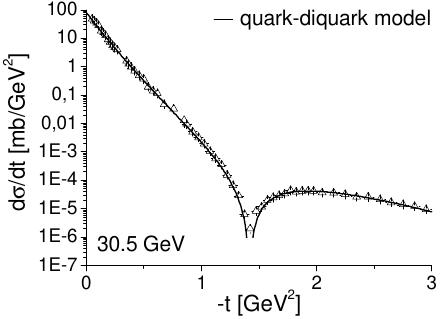}
\end{center}
\caption{Comparison of the experimental differential cross section with the
results of the quark-diquark model. Data at $\sqrt{s}=30.5$ GeV are taken from
\cite{elastic}.}%
\label{Fig_31}%
\end{figure}

Data at all ISR energies \cite{elastic,p(0)} were analyzed. It turns out that
the model works very well indeed which is by no means a trivial
conclusion\footnote{For instance, an analogous calculation performed in the
model with the assumption that the proton consists of three uncorrelated
constituent quarks led to negative conclusion \cite{pp-3q}.}. One example of
such calculation is shown in Fig. \ref{Fig_31}, where the differential cross
section $d\sigma/dt$ at the ISR energy $30.5$ GeV, evaluated from the model,
is compared with experimental data \cite{elastic}. One sees a rather
impressive agreement. A detailed discussion goes beyond the scope of this
paper and will be given elsewhere \cite{ab-ab}. Here we are interested only in
the resulting values for $\sigma_{qq}/\sigma_{NN}$ and $w_{N}^{(q+d)}$ which
are necessary for evaluation of the R.H.S. of (\ref{start}).

From the point of view of the present investigation, the most important
observation is that both the ratio $\sigma_{qq}/\sigma_{NN}$ and the average
number of wounded constituents in nucleon-nucleon collisions $w_{N}^{(q+d)}$
seem almost entirely independent of the details of the model (provided that,
as explained above, the parameters are adjusted to describe correctly the
proton-proton elastic data). The obtained values are:
\begin{equation}
%\sigma_{qq}/\sigma_{NN}=0.147\div0.148;\;\;\;w_{N}^{(q+d)}=w_{q}+w_{d}=1.182\div1.186.
\sigma_{qq}/\sigma_{NN}=0.147 - 0.148;\;\;\;w_{N}^{(q+d)}=w_{q}+w_{d}=1.182 - 1.186.
\end{equation}
These values\footnote{Note that $\sigma_{qq}/\sigma_{NN}>1/9$, indicating
presence of shadowing.}, supplemented by the relation (\ref{124}), are used
for evaluation of the R.H.S. of (\ref{start}).

\section{Au+Au collisions}

Having determined the parameters of the model from the elastic pp data, we
could evaluate its predictions for the particle production in Au-Au collisions
which is the main goal of this investigation.

\vskip      0.2cm\begin{figure}[h]
\begin{center}
\includegraphics[scale=1.1]{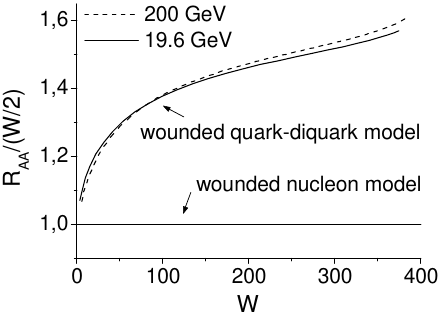}
\end{center}
\caption{{Predictions of the wounded quark-diquark model (for }$W\geq5${)
compared with those from the wounded nucleon model.}}%
\label{Fig_R}%
\end{figure}

Since the PHOBOS data are presented versus the number of the wounded nucleons
in both colliding nuclei ($2w_{A}^{(N)}=W$), we have to calculate also
$w_{A}^{(N)}$ as a function of the impact parameter of the collision. This can
be obtained from the well-known formula \cite{wnm}:
\begin{equation}
w_{A}^{(N)}(b)=\frac{A}{\sigma_{AA}(b)}\int T(b-s)\left\{  1-[1-\sigma
_{NN}T_{NN}(s)]^{A}\right\}  d^{2}s,
\end{equation}
with $\sigma_{NN}(s)$ in a Gaussian form with $\sigma_{NN}(0)=0.92$ taken from
our estimates (it agrees very well with the data \cite{p(0)}).

For the nuclear density we have been using the standard Woods-Saxon formula
with the nuclear radius $R_{Au}=6.37$ fm, and $d=0.54$ fm.

In Fig. \ref{Fig_R} the predicted ratio $R_{AuAu}/w_{Au}^{(N)}$ (which shows
explicitly the deviation of our model from the traditional wounded nucleon
model) is plotted versus $2w_{Au}^{(N)}=W$ for $W\geq5$. One sees that the
model explains naturally the increase of the production multiplicity from one
wounded nucleon with increasing centrality of the collision.

\begin{figure}[h]
\begin{center}
\includegraphics[scale=0.97]{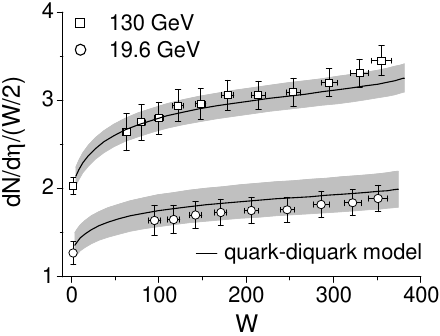}  \hspace{0.2cm}
\includegraphics[scale=0.97]{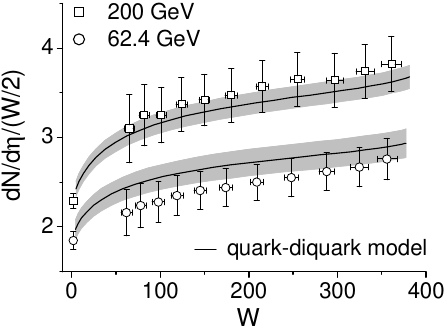}
\end{center}
\caption{The predictions of the wounded quark-diquark model (for $W\geq5$)
compared with the data from PHOBOS coll. \cite{AuAu-130,AuAu-19-200,AuAu-62}.
The shaded areas reflect the inaccuracy in the pp data. }%
\label{AA_Gauss_fig}%
\end{figure}

The comparison of the model with the PHOBOS data on particle production per
one wounded nucleon is shown in Fig. \ref{AA_Gauss_fig}.

\vskip      0.2cm

The data on particle production in pp collisions, necessary to obtain the
model predictions, were taken from UA5 collaboration, as quoted in
\cite{AuAu-19-200,AuAu-62}\footnote{For $\sqrt{s}=62.4$ and $200$ GeV they are
taken directly from UA5 data, for $\sqrt{s}=19.6$ and $130$ GeV they are
interpolated.}. They are also shown in the Fig. \ref{AA_Gauss_fig} as points
at $W=2$. This introduces some uncertainty, as indicated by the shadowed
areas\footnote{Unfortunately, the pp data from PHOBOS coll. are still not
available.}.

The inelastic proton-proton cross sections needed for this calculations were
taken as $\sigma_{NN}=32$ mb, $36$ mb, $41$ mb and $42$ mb at energies
$\sqrt{s}=19.6,$ $62.4,$ $130$ and $200$ GeV, respectively. Our model then
gives the following values of the ratio of the integrated inelastic
quark-quark to proton-proton cross sections: $\sigma_{qq}/\sigma_{NN}=0.147,$
$0.148,$ $0.148,$ $0.149$. Finally, the number of wounded quarks and diquarks
in a single proton-proton collision $w_{q}+w_{d}=1.183,$ $1.185,$ $1.186,$ $1.187.$

One sees that, within the experimental accuracies a very good agreement both
in shape and in absolute value is obtained.

\section{Comments and discussion}

Several comments are in order.

(i) The concept of wounded nucleons and of wounded constituents is based on
two ideas: (a) during the interaction, the ''wounded'' object acts as one unit
and (b) particle emission process takes much longer time than it is needed for
the projectile to pass the internuclear distance. These assumptions can be
qualitatively justified only for \textit{soft} collisions where the momentum
transfer and transverse mass of the emitted partons are small enough.

The first condition requires that the momentum transferred to the projectile
is smaller than its inverse size. For the size of order of $1$ fm, this limits
the momentum transfer to about $200$ MeV.

The second condition demands that the emission time:
\begin{equation}
t\sim\frac{\gamma}{m_{\perp}}\approx\frac{e^{y}}{2m_{\perp}},
\end{equation}
where $\gamma$ is the Lorentz factor of the emitted particle (parton) in the
rest frame of the target nucleus and $y$ is its rapidity in the same frame,
should be significantly greater than the intranuclear distance. For $y>2$ this
limits the transverse mass of the emitted partons below $200$ MeV. Of course
for the observed final hadrons this limit may be significantly higher.

These estimates are, surely, rather crude. A more detailed verification of the
model for particles with various masses and transverse momenta will be of
great interest, as it may help to understand better the very concept of a
wounded constituent.

(ii) As seen from these arguments, the model is not expected to apply in the
fragmentation region of the projectile and target where, moreover, important
energy-momentum conservation effects, as well as secondary interactions inside
the nucleus must be present. Therefore we focus our attention on the central
rapidity region which, at RHIC energies, is well separated from the
fragmentation regions.

(iii) The model assumes that a wounded quark produces the same number of
secondary partons as a wounded diquark. This is not unreasonable since the
colour content of both constituents is the same ($3$ and $\bar{3}$) and that,
probably, the colour charge of the projectile is the main factor determining
the emission intensity. We admit, however, that since the actual dynamics of
the soft production process is not yet understood, this argument can be
questioned. The good agreement of the model with data, as presented in this
paper, may thus serve as an (indirect) confirmation of the important role of
colour dynamics in the process of particle production.

(iv) The predictions of the model described in this paper refer to the
emission of ''primary'' partons and do not take into account further
development of the system during its expansion and final formation of hadrons.
The observed agreement with data indicates that the space-time development of
the system, despite presence of the well-known collective effects, does not
introduce drastic changes in its global characteristics. This may be
considered as an argument in favour of the laminar hydrodynamic expansion
suggested already by MC simulations of this process which seems to indicate
very small viscosity of the created medium \cite{hydro}.\footnote{This
argument is due to W. Czyz (private communication).}

(v) We have verified that the main properties of the quark-diquark picture of
the nucleon obtained from analysis of the pp elastic data are not sensitive to
the details of the calculations (several forms of the distribution of
constituents and of their cross-sections were considered). The typical values
of the parameters are $R_{q}\approx0.3$ fm, $R_{d}\approx0.75$ fm,
$R\approx0.3$ fm, $A_{dd}\approx0.55$, $A_{qd}\approx0.5$, $A_{qq}\approx
1$.\footnote{$A_{qq}$ depends somewhat on the value of $\lambda$ which is not
very well determined. For $\lambda$ not far from $1/2$, $A_{qq}=1$.} Thus in
our model the diquark appears to be rather large, comparable to the size of
the proton. It is remarkable that this feature agrees well with other
estimates \cite{Wilczek}, based on rather different arguments.

(vi) In this paper we have only discussed the symmetric Au-Au collisions. It
would be also interesting to check the model for asymmetric collisions. We
have verified that in case of d-Au collisions at $y=0$ the formula
(\ref{start'}) gives the result which does not differ very much (less than
$10$ \%) from that of the wounded nucleon model. Thus the observed good
agreement of the wounded nucleon model \cite{wnm-deuteron} with the PHOBOS
data on d-Au collisions \cite{data-dAu} is not destroyed in our approach.

In conclusion, we have formulated a model in which the soft collisions of the
nucleon are described in terms of interactions of its two constituents: a
quark and a diquark. The model can be adjusted to describe very precisely the
elastic proton-proton scattering data. Supplemented with the idea of wounded
constituents, the model accounts rather well for the centrality dependence of
particle production in the central rapidity region at RHIC energies.

\bigskip

\textbf{Acknowledgements}

We have greatly profited from discussion with Wieslaw Czyz to whom we also
like to thank for a critical reading of the manuscript. Discussions with B.
Wosiek, M. Praszalowicz and correspondence with G. Ripka are also highly appreciated.
This investigation was partly supported by the MEiN research grant 1 P03B 045
29 (2005-2008) and the Polish Science Foundation (FNP)
scholarship for the year 2006.

\end{document}